\documentclass[preprint2]{aastex}
\usepackage{picture}

\newcommand{\gtabouteq}{\,\hbox{\raise 0.5 ex \hbox{$>$}\kern-.77em 
                    \lower 0.5 ex \hbox{$\sim$}$\,$}}       
\newcommand{\ltabouteq}{\,\hbox{\raise 0.5 ex \hbox{$<$}\kern-.77em 
                     \lower 0.5 ex \hbox{$\sim$}$\,$}}
\newcommand{\hb}{\hfill\break}

\slugcomment{To appear in the Astronomical Journal}

\shorttitle{CHANG-ES: I}
\shortauthors{Irwin et al.}

\begin{document}


\title{Continuum Halos in Nearby Galaxies -- an EVLA Survey \\
    (CHANG-ES) -- I: Introduction to the Survey}


\author{Judith Irwin\altaffilmark{1}, Rainer Beck\altaffilmark{2},
R. A. Benjamin\altaffilmark{3}, Ralf-J{\"u}rgen Dettmar\altaffilmark{4},
Jayanne English\altaffilmark{5},\\
George Heald\altaffilmark{6}, Richard N. Henriksen\altaffilmark{7},
Megan Johnson\altaffilmark{8},
Marita Krause\altaffilmark{9}, Jiang-Tao Li\altaffilmark{10},
Arpad Miskolczi\altaffilmark{11},
Silvia Carolina Mora\altaffilmark{12},
E. J. Murphy\altaffilmark{13}, Tom Oosterloo\altaffilmark{14},\\
Troy A. Porter\altaffilmark{15}, Richard J. Rand\altaffilmark{16},
D. J. Saikia\altaffilmark{17},
Philip Schmidt\altaffilmark{18},\\
A. W. Strong\altaffilmark{19}, Rene Walterbos\altaffilmark{20},
Q. Daniel Wang\altaffilmark{21}}
\and
\author{Theresa Wiegert\altaffilmark{22}}

\altaffiltext{1}{Dept. of Physics, Engineering Physics \& Astronomy, 
Queen's University, Kingston, ON, Canada, K7L 3N6, {\tt irwin@astro.queensu.ca}.}
\altaffiltext{2}{Max-Planck-Institut f{\"u}r Radioastronomie, Auf dem H{\"u}gel 69,
53121, Bonn, Germany,
{\tt rbeck@mpifr-bonn.mpg.de}.}
\altaffiltext{3}{Dept. of Physics, University of Wisconsin at Whitewater, 800 West
Main St., Whitewater, WI, USA, 53190,
{\tt benjamin@wisp.physics.wisc.edu}.}
\altaffiltext{4}{Astronomisches Institut, Ruhr-Universit{\"a}t Bochum, 44780 Bochum,
 Germany,
{\tt dettmar@astro.rub.de}.}
\altaffiltext{5}{Department of Physics and Astronomy, 
University of Manitoba, Winnipeg, Manitoba, Canada, R3T 2N2,
{\tt jayanne\_english@umanitoba.ca}.}
\altaffiltext{6}{Netherlands Institute for Radio Astronomy (ASTRON), 
Postbus 2, 7990 AA, Dwingeloo, The Netherlands,
{\tt heald@astron.nl}.}
\altaffiltext{7}{Dept. of Physics, Engineering Physics \& Astronomy, 
Queen's University, Kingston, ON, Canada, K7L 3N6, {\tt henriksn@astro.queensu.ca}.}
\altaffiltext{8}{National Radio Astronomy Observatory, P. O. Box 2, Greenbank, WV, USA, 24944,
{\tt mjohnson@nrao.edu}.}
\altaffiltext{9}{Max-Planck-Institut f{\"u}r Radioastronomie,  Auf dem H{\"u}gel 69,
53121, Bonn, Germany,
{\tt mkrause@mpifr-bonn.mpg.de}.} 
\altaffiltext{10}{Dept. of Astronomy, University of Massachusetts, 710 North
Pleasant St., Amherst, MA, 01003, USA, 
{\tt jiangtao@astro.umass.edu}.} 
\altaffiltext{11}{Astronomisches Institut, Ruhr-Universit{\"a}t Bochum, 44780 Bochum, Germany, 
{\tt miskolczi@astro.rub.de}.}
\altaffiltext{12}{Max-Planck-Institut f{\"u}r Radioastronomie,  Auf dem H{\"u}gel 69,
53121, Bonn, Germany,
{\tt cmora@mpifr-bonn.mpg.de}.} 
\altaffiltext{13}{Observatories of the Carnegie 
Institution for Science, 813 Santa Barbara Street, Pasadena, CA, 91101,
USA,  {\tt 
emurphy@obs.carnegiescience.edu}.} 
\altaffiltext{14}{Netherlands Institute for Radio Astronomy (ASTRON), Postbus 2,
7990 AA, Dwingeloo, 
The Netherlands, {\tt oosterloo@astron.nl}.} 
\altaffiltext{15}{Hansen Experimental Physics Laboratory, Stanford University, 
452 Lomita Mall, Stanford, CA, 94305, USA, {\tt tporter@stanford.edu}.}
\altaffiltext{16}{Dept. of Physics and Astronomy, University of New Mexico, 
800 Yale Boulevard, NE, Albuquerque, NM, 87131, USA, {\tt rjr@phys.unm.edu}.} 
\altaffiltext{17}{National Centre for Radio Astrophysics, 
TIFR, Pune University Campus, Post Bag 3, Pune, 411 007, India,
 {\tt djs@ncra.tifr.res.in}.}
\altaffiltext{18}{Max-Planck-Institut f{\"u}r Radioastronomie,  Auf dem H{\"u}gel 69,
53121, Bonn, Germany,
{\tt pschmidt@mpifr-bonn.mpg.de}.}
\altaffiltext{19}{Max-Planck-Institut f{\"u}r extraterrestrische Physik, 
Garching bei M{\"u}nchen, Germany, {\tt aws@mpe.mpg.de}.}
\altaffiltext{20}{Dept. of Astronomy, New Mexico State University, 
PO Box 30001, MSC 4500, Las Cruces, NM 88003, USA, {\tt rwalterb@nmsu.edu}.}
\altaffiltext{21}{Dept. of Astronomy, University of Massachusetts, 710 North
Pleasant St., Amherst, MA, 01003, USA, 
{\tt wqd@astro.umass.edu}.}
\altaffiltext{22}{Dept. of Physics, Engineering Physics \& Astronomy,
Queen's University, Kingston, ON, Canada, K7L 2T3, 
{\tt twiegert@astro.queensu.ca}.}


\begin{abstract}
We introduce a new survey to map the radio continuum halos
of a sample of 35 edge-on spiral galaxies at 1.5 GHz and
6 GHz in all polarization products.  The survey is exploiting the new wide bandwidth
capabilities of the Karl G. Jansky Very Large Array (i.e. the Expanded Very Large Array, or EVLA)
in a variety of array configurations (B, C, and D)
in order to compile the most comprehensive data set yet obtained for
the study of radio halo properties. This is the first survey of radio halos to include
all polarization products.

In this first paper, we outline the scientific motivation of the survey, 
the specific science goals, and the expected
 improvements in noise levels and spatial coverage from the survey.
Our goals include investigating the physical conditions and origin of halos,
characterizing cosmic ray transport and wind speed, measuring Faraday rotation
and mapping the magnetic field, probing the in-disk and
extraplanar far-infrared - radio continuum
relation, and reconciling non-thermal radio emission with high-energy gamma-ray
models.   The sample size allows us to search for correlations between
radio halos and other properties, including environment, star formation rate,
and the presence of AGNs.  In a companion
paper (Paper II) we outline the data reduction steps and present the
first results of the survey for the galaxy,
NGC~4631.

\end{abstract}


\keywords{ISM: bubbles -- (ISM:) cosmic rays -- ISM: magnetic fields --
galaxies: individual (NGC~4631) -- galaxies: magnetic fields -- radio continuum: galaxies}



\section{Introduction}
\label{sec:introduction}

\begin{figure*}[b]
\setlength{\unitlength}{0.1in}
\begin{picture}(75,60)
\put(5,0){\includegraphics{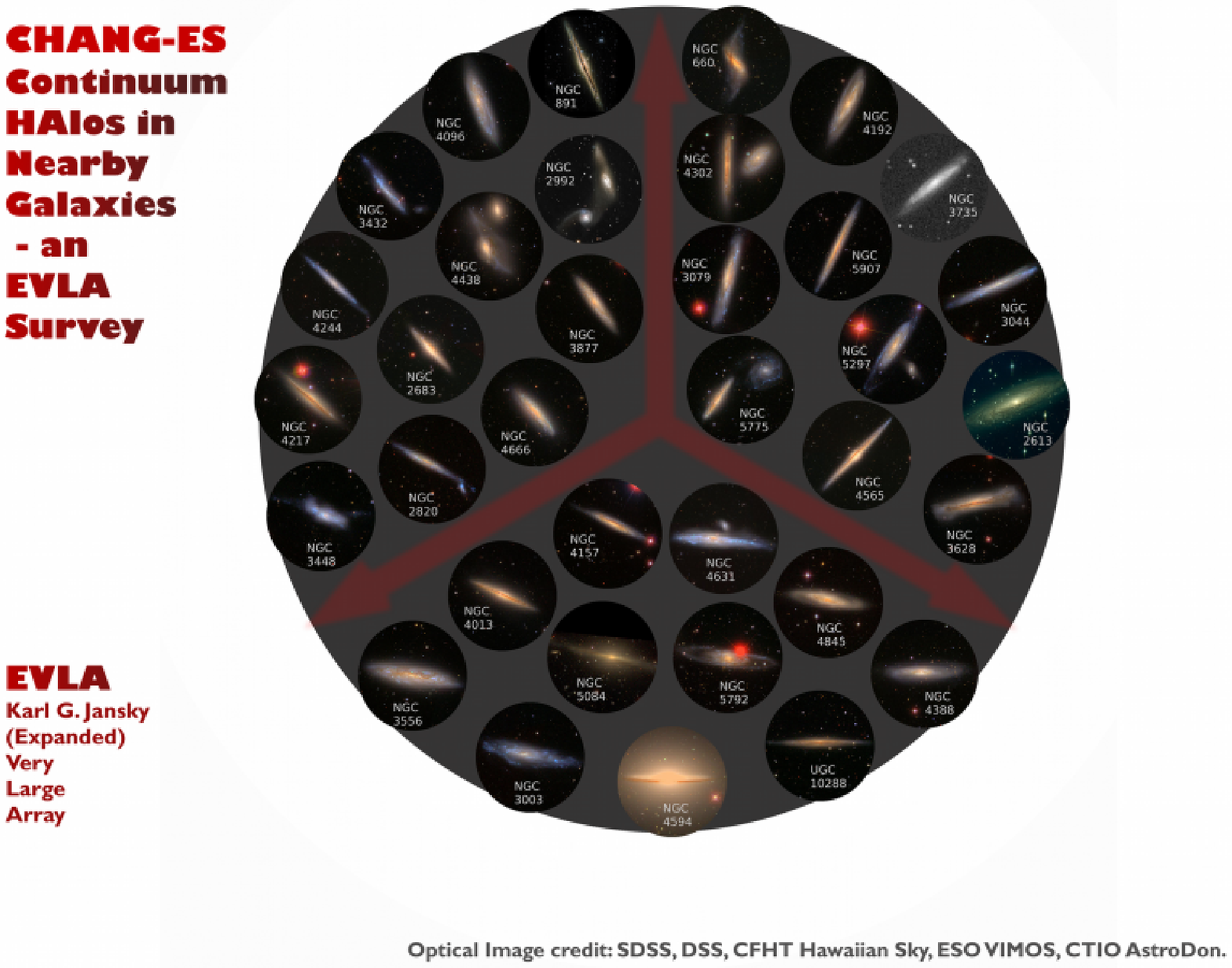}}
\end{picture}
\caption{The CHANG-ES logo, showing optical images of the galaxy sample and the `Y' shape of the EVLA.}
\label{fig:logo}
\end{figure*}

We introduce a new survey to detect radio continuum
halos\footnote{By `halo', we mean the 
the gas, dust, cosmic ray, and magnetic field components above a galaxy's plane, rather than
stellar or dark matter halos. We will
refer to
the region, $0.2\,\ltabouteq\,z\,\ltabouteq\,1$ kpc, as the
disk-halo interface and use `halo' for emission on
larger scales ($z\,\gtabouteq\,1$ kpc).  `High-latitude' or `extraplanar' are also used to describe either
of these components. }
in spiral galaxies using the Karl G. Jansky Very Large Array (hereafter, the 
Expanded Very Large Array, or EVLA).
``Continuum Halos in Nearby Galaxies -- an EVLA Survey'' (CHANG-ES)
is observing 35 nearby edge-on galaxies (Table~\ref{table:sample}, Fig.~\ref{fig:logo}) in the radio continuum 
at two frequencies (1.5 GHz and 6 GHz, i.e. in L-band and C-band, respectively) 
in all polarization products,
and over a range of spatial scales.  Our overarching goals are to characterize the
nature and prevalence of radio halos, their magnetic fields and the cosmic
rays that illuminate those fields. By initiating a survey, 
we can address the connection between radio halos
and the underlying galaxy disk as well as galaxy environment 
and provide a coherent data set with 
legacy value for the scientific community.

 In this first paper, our goals are to outline the motivation for
the survey (Sect.~\ref{sec:motivation}), to describe its
scientific goals 
 (Sect.~\ref{sec:science}), to introduce the galaxy sample and its characteristics
 (Sect.~\ref{sec:sample}), and to detail the EVLA observations, including our specific
goals for sensitivity, resolution and spatial coverage
(Sect.~\ref{sec:observations}).  In a companion paper \citep[][hereafter, Paper II]{irw12b}
we present the first results of the survey from test observations of the galaxy, NGC~4631.

\section{Motivation}
\label{sec:motivation}
Since the early pioneering observations of edge-on galaxies 
\citep{eke77,all78,hum89,hum91},
 it has been clear that radio continuum emission
can extend well above galaxy disks, that is, 
 to $z$ heights\footnote{The height, 
$z$, is measured perpendicular to the plane of a galaxy disk.} greater
than 500 pc and often to many kpc.  Radio continuum emission represents
just one component of the multiphase gaseous halos that are typically present.
For example, halos have been observed in H$\alpha$ emission and other
 optical lines
(called the `extraplanar diffuse ionized gas' or eDIG), 
in soft X-ray emission,
in the 21 cm HI line, in infra-red (IR) emission from dust, and 
in molecular species such as CO and PAHs
\citep[e.g.][]{bre97,lee01,irw90,mar01,hee09a,hee09b,wha09,wan95,wan01,how99,ros03,ran96a}.
The halo radio continuum emission is primarily from synchrotron radiation
\citep[e.g.][]{irw99}
implying the presence of cosmic rays (CRs) and magnetic fields out of the
plane.

Although radio continuum studies of edge-on galaxies
 have been conducted in the past
\citep{hum89,hum91,dum95,irw99,irw00},
the limited number of galaxies for some of the surveys, frequency coverage, sensitivity limits, and/or
ability to detect a variety of spatial scales have made it difficult to 
systematically and consistently characterize
the properties of radio halos. 
For example, those studies have observed (in the same order as referenced above) 4 galaxies at
1.5 GHz, 181 galaxies at 5 GHz, 6 galaxies at 10.7 GHz, 16 galaxies at 1.4 and 6 GHz,
and 16 galaxies at 1.4 GHz.  The spatial scales that were detected as well as the sensitivity
limits (see below) varied significantly. 
CHANG-ES is attempting to improve upon previous surveys in a number of ways.

To begin with, 
a galaxy sample of significant size (Sect.~\ref{sec:sample}) is being observed over a variety of 
angular spatial scales (Sect.~\ref{sec:observations}). The latter is 
particularly important since
previous observations of sufficient spatial resolution and 
sensitivity have shown that structure exists in halos on virtually any scale
that is observed.  The
`smooth' halos that are observed on broad scales resolve into
discrete features on a variety of smaller scales. 

Low frequencies (1.5 and 6 GHz) have been chosen 
for their sensitivity to
non-thermal emission 
and for the prospects of carrying out Faraday rotation analyses in galaxy
disks and halos.  Ours will be the first systematic survey for Faraday rotation in 
galaxy halos.  The frequency choice also ensures that complementary data (for example, single dish data)
can be obtained, where
needed.  

CHANG-ES can also now exploit
the wide bandwidths that are provided by
 the EVLA's Wideband Interferometric Digital ARchitecture (WIDAR) correlator. 
These wide bandwidths permit unprecedented improvements in 
continuum sensitivity.
For example,
 the 5 GHz survey of \citet{hum91}
achieved a best rms noise of 80 $\mu$Jy beam$^{-1}$ while the 1.4 GHz and 5 GHz surveys of
\citet{irw99} achieved best rms values of 43 $\mu$Jy beam$^{-1}$ and 46 $\mu$Jy beam$^{-1}$,
 respectively.  
Our sensitivity limits
should improve upon these previous surveys, in some cases by more than an order of magnitude
(see Sect.~\ref{sec:observations}).

In addition, the wide bands provide a wealth of spectral information which will be unique to this survey.
For example, it is now possible to measure a spectral index {\it within} a specific observing
band
(and possibly more than one for strong emission if there is curvature).
  We will therefore obtain at least three spectral indices from observations at two
frequencies (Sect.~\ref{sec:separation}).



In summary, CHANG-ES improves on previous surveys by observing a large number of
galaxies over a variety of well-defined angular spatial scales at two frequencies, and to low noise
limits.
 By exploiting the wide-band capabilities of the EVLA, a wealth of spectral index information
is available.  Moreover,  CHANG-ES is the first survey of
edge-on galaxies for which all polarization products will be obtained, providing 
important information
on halo magnetic fields.


\section{Science Goals}
\label{sec:science}

\subsection{Physical Conditions and Origin of Gaseous Halos}
\label{subsec:physical_conditions}
The disk-halo interface and broader halo are dynamic entities that connect starkly different
environments:  the normal ISM of a galaxy disk, and the rarefied intergalactic medium (IGM).
 A
variety of observational evidence \citep[e.g.][]{dah95,ros03,str04,ros08,tul06} 
  points
to a connection between the supernova rate per unit area in the disk and the intensity of
the halo radio continuum, X-ray and eDIG emission.  In a few cases, it is also
possible to associate discrete disk-halo features with underlying star forming regions, the
latter also presumably being a source of eDIG ionization. 

Most models have therefore
 focussed on the critical link between 
extra-planar gas and
star formation-related activity (largely supernovae) in the underlying disk. 
Such models have included galactic
chimneys \citep{nor89}, fountains \citep{sha76}, superbubble blowout \citep{mac99}
and CR-driven Parker
instabilities 
\citep[e.g.][]{bre91,han02}.  
  
CRs, in particular, appear to be an essential ingredient in driving outflows
\citep[e.g.][]{ipa75,bou90,eve08,eve10} and it is this component, along with
magnetic fields, that are critical to understanding outflow physics.  In the disk,
for example, CRs, the magnetic field, and hot gas have roughly equal pressures, whereas in
the halo far from the plane, the 
CR and magnetic field pressures appear to dominate; however, it is not clear
whether such statements are universally true.  
Although we now have information on the strength and configuration of halo magnetic 
fields in some galaxies [for reviews, see \citet{kra09} and \citet{kra11}], an important
goal of CHANG-ES is to consistently measure 
non-thermal pressures in galaxies with a variety of SFRs.  Discrete disk-halo
features can also be examined for possible associations with star forming sites in the disk.

Yet, in spite of the abundant evidence that links star formation to extra-planar gas, 
there are still some puzzles that suggest that underlying star formation is not a sufficient
explanation for the existence of halos.
There is evidence, for example, that
a galaxy's global magnetic field configuration seems not to be influenced
by SFR \citep{kra09}.
 Optical line ratios suggest that a source of heating other than photoionization
from in-disk sources is required to ionize the eDIG
\citep[e.g.][]{tul00,ran11}; possibilities include shocks, and
 heating from magnetic reconnection \citep{zim97}.
In the latter case, for example, 
specific magnetic field topologies are predicted \citep{han02}. In another model
\citep{gis09} vertical inflow plays a role.  In still another, hydraulic jumps at the 
arm-interarm boundary produce long-lived vertical features
\citep{mar98}.
SFR also does {\it not} seem to affect radio continuum scale heights \citep{kra09,kra11}.

In addition, evidence for a direct relationship
 between HI supershells and SFRs is weaker.  
Expansion energies of some HI supershells, where measurable, appear to be too high
to be accounted for assuming reasonable numbers of supernovae and stellar winds
\citep[][and references therein]{spe04,sil06} and some HI supershells appear to be present
outside of the star forming disks of galaxies \citep{irw90,cha01}, prompting suggestions of other
sources of energy (but see \citet{cha11})
 or that the shells have been formed from
 impacting external clouds \citep[e.g][]{ran96b}.  

The fact that 
halos {\it lag} in comparison to the underlying disk
\citep{swa97,tul00,sch00,lee01,fra04,hea07},
and the difficulty in explaining these lags via dynamical/ballistic models with
the gas originating in the disk
\citep{fra06,hea07} 
have led to interpretations that include the accretion of external material onto
galaxy disks 
\citep{oos07,fra07}. We need to understand the effects of
magnetic field pressure and tension on the dynamics of cycling gas \citep{ben02}
in order to account for energies and velocity gradients in such systems.

It is clear that  
circulation and/or outflow can have a dramatic effect on
the evolution of the galaxy itself, from modifications of the metallicity gradient  
to understanding the energy and mass budgets of galaxies, to altered SFRs.
Mass flux estimates thus far \citep{bre94,wan95,fra02}  
 imply that disk-halo outflows
are responsible for moving a large amount of gas
and may even result in extraplanar star formation \citep[e.g.][]{tul03}.
 That a disk-halo flow is responsible
for some of the Galactic intermediate-velocity clouds (IVCs) is also
likely \citep[e.g.][]{wak97}.

Galactic 'feedback' is an important component in $\Lambda$CDM numerical models
of galaxy formation.
Whether feedback is in the form of central starbursts, 
global starbursts, 
or active galactic nuclei 
(AGNs) is not yet known \citep[for a review, see][]{bau06}, but
sometimes type Ia supernovae (objects that are normally associated with Population II objects within
a galaxy), are implicated as an important
driver \citep[][and references therein]{lij11,wan10}. 
Observations of S0 galaxies and bulges, for example, indicate that observed X-ray halos are weaker 
than feedback would suggest \citep[e.g.][]{lij11}.
Moreover
some galaxies with {\it both} weak 
AGN and star formation can also
have soft X-ray luminous halos \citep[e.g.][]{liz11}.
But the relations of such outflows to the stellar mass and to the galactic
environment and evolution history remains uncertain.
While radio continuum emission 
appears to be mainly related to Population I objects as described above, it is important
to verify whether this is always true and to search for deviations that may reveal
a link to older components.

\subsection{Cosmic Ray Transport and Wind Speed} 
\label{subsec:cr-transport} 
Theoretical and observational
effort on the transport mechanism for cosmic rays (diffusive, convective or both)
has been carried out for some time
\citep[e.g.][]{str78,poh90,dur98}.
Cosmic ray electrons (CREs) can lose energy via a variety of processes, i.e.
synchrotron, Inverse Compton,
adiabatic, ionization, non-thermal Bremsstrahlung, although the latter two
can be neglected in normal galaxies \citep{con92}.  The relative importance of
the remainder depends on environment, but \citet{hee09a} have shown that
the shortest timescales are associated with
 synchrotron and adiabatic losses in the outflow galaxy, NGC~253.
An observational consequence (and one that CHANG-ES can look for) 
is  a measurable steepening of the spectral index 
where
adiabatic losses transition to radiative losses (see Sect.~\ref{sec:separation} for
a discussion of the separation of thermal from non-thermal emission).  An analysis of 
the variation of
magnetic field strength and
radio spectral index with $z$ height,
($B(z)$ and $\alpha (z)$, respectively) 
can also provide a wealth of information
on CR transport. 

Outflow velocity is an important datum for input to
numerical models of galaxy formation.   
Since winds are thought to be driven by CRs and couple to
the thermal outflow, a measurement of CRE bulk velocity should apply to the thermal wind
speed as well. 
The CRE bulk speed can be determined from the radio scaleheight at several frequencies
and the CRE lifetime \citep{hee09a}. 
Applying this method to more galaxies and in more detail is an important
goal of CHANG-ES.
The variation of these quantities with position
across the disk can also constrain outflow models and provide information on CR
transport within the disk as well.
 
Outflows from galaxies may also be
the mechanism whereby the intergalactic medium became magnetized \citep{kro99}.
The bulk CRE outflow speed in NGC~253, for example, achieves
 300 km s$^{-1}$ in some locations, i.e. greater
than the escape velocity, implying that this outflow should enrich the intergalactic medium
\citep{hee09a}.
Although the conditions in nearby galaxies may differ substantially from the high redshift
universe, we can address very specific issues, such as how outflows scale with SFR,
SFR/area, the presence of compact nucleii, and gravitational potential.


\subsection{Faraday Rotation and the Origin of Galactic Magnetic Fields}
\label{subsec:faraday}
A fundamental question which CHANG-ES may help to address is the origin of the field itself.
  Seed fields
alone should wind up too quickly to account for
the polarization angles that are observed
in galaxies.  The $\alpha-\Omega$ dynamo 
 is thought to be the best alternative for generating observed
fields \citep{bec96}.  Although the mean-field dynamo might operate separately
in the disk and halo regimes, the regime with the more active dynamo may `enslave' the other,
resulting in a coupling between disk and halo \citep{mos08,mos10}. 

 Yet in the few cases where halo fields are detected,
they can show geometries that are not easily explained by dynamos alone. 
Indeed, the observational evidence for regular 
 (i.e. dynamo-generated) fields in galaxy halos so far is rather weak -- possibly
NGC253 \citep{hee09b} and indirect evidence from polarized intensity asymmetries seen
at 20 cm in 7 galaxies \citep{bra10}.  Faraday rotation studies of our own
Galaxy also show no coherent vertical field structure at the Sun's position \citep{mao10}.

Rather, the observations appear to support dynamo-generated fields 
in the disk that couple to outflowing winds
\citep[see][]{kra09}.
For example, in NGC~4631, the field appears to be perpendicular to the disk where star formation is larger, but
at larger radii where the SFR declines, the field becomes parallel to the disk again \citep{kra04}.
`X-shaped' magnetic structures that have been observed in edge-on galaxies have been reproduced by 
\citet{han09} in global MHD simulations of CR-driven dynamos.  In addition, magnetic helicity
conservation requires that outflows be present \citep{shu06,sur07}.
 The need for winds in explaining
magnetic field structures
is certainly consistent with the abundant evidence for outflows that has been outlined
above.

Understanding the structure of the magnetic field requires observations of
Faraday rotation, which measures the strength of the ordered magnetic field along the line
of sight, and synchrotron polarization, which measures the ordered magnetic field in the plane of the
sky.  
\citet{soi11} provide an example of the application of rotation measure
information to the study of an edge-on galaxy with a radio halo. 
  By utilizing many frequency channels over two frequency
bands
and obtaining polarization data over a variety of spatial scales, CHANG-ES observations can
disentangle these effects and (with models) even start to probe the 3-dimensional structure of 
the magnetic field.   
Rotation measure (RM) synthesis techniques described in 
\citet{bre05} and \citet{hea09},
 which make use of the wide frequency
response and many channels of the WIDAR correlator, are being employed.

\subsection{The FIR-Radio Continuum Correlation}
\label{subsec:fir-radio}
Young hot massive stars become supernovae which are believed to be the main sites of CR acceleration 
in galaxy disks, and hot massive stars are also important sources of photons that can heat the
dust in disks giving rise to FIR emission.  This common origin is thought to be at the heart of
the FIR-radio continuum relation which has been observed among and within galaxies 
\citep[e.g.][]{hel85,mur06}.

Since the mean free path of a dust-heating photon is
smaller than the diffusion length of a CR electron, a radio image of a galaxy will be
a smoothed version of a FIR image. 
Retrieval
of scale length information can provide important information on the star formation timescale and radiation intensity
\citep[see][]{mur08} and on the degree of field ordering \citep{tab12}.
Moreover, by separating thermal from non-thermal radio emission
(Sect.~\ref{sec:separation}) and correlating the result
with IR emission in different wavebands, we can investigate the
warm dust-thermal radio correlation and the cold dust-nonthermal radio correlation
\citep[e.g.][]{bra03,tab07}.
 
In halos far from sites of in-disk SF, one would expect the difference between
mean free paths to be greater still, and that the FIR-radio relation may even break down
altogether at some scale height.   
  Together with
IR images, CHANG-ES should be able to investigate to what extent the FIR-radio correlation
is or is not valid in galaxy halos.  Environment may also play an important part. 
The low density relativistic plasma component of galaxy halos can provide a 
sensitive tracer to environmental effects. 
 \citet{mur09},
for example, have shown that the relation is modified 
in Virgo Cluster spirals, likely because of galaxy motions with respect to the
intra-cluster medium. The CHANG-ES sample includes galaxies within a wide range of
environments.

\subsection{CRs and High Energy Modeling}
\label{subsec:high-energy}

Synchrotron emission results from the electron component of CRs and
so there is an tight connection between the gamma-ray emission that
is produced by CR nuclei and lepton interactions with gas and radiation fields, and the radio continuum emission
that CHANG-ES will observe.
The
GALPROP code\footnote{http://galprop.stanford.edu/}  
\citep{mos98,mos02,por08,str00,str04,str09,str10}
 self-consistently links these and other
components in a model of the Galaxy, predicting (polarized and total) broad-band fluxes from radio to gamma-ray
energies; it also allows for the inference of other parameters such as the escape probability of particles and the
diffusion coefficient.

With modifications, this model will be applied to other galaxies to predict synchrotron scale heights together 
with
gamma-ray intensities from each galaxy.
Gamma rays give independent information on the energy density of the CRs and
hence allow the determination of magnetic field strengths without the assumption of energy equipartition. With the
ability to put strict observational limits on galaxy radio halo sizes (one of the largest sources of uncertainty in
GALPROP models of the Milky Way) we will be able to constrain the CR diffusion and energy losses for all species
(primary/secondary electrons and nuclei). At present, NGC~253 and M82 are the only edge-on galaxies that have been
detected in high-energy gamma rays  \citep{ace09,abd10,acc09}, but our gamma-ray flux 
predictions can be constrained by current upper
limits.  

 X-ray emission generated from
inverse-Compton scattering also provides important relevant data.
 For example,
 existing X-ray observations from Chandra and
XMM-Newton may already be useful for measuring or tightly constraining the hard X-ray (e.g., 2-10 keV) intensity
and its spatial distributions in galactic halos. The combination of the X-ray and the radio measurements can be
used to measure the magnetic field and CRe density.
Future X-ray satellites\footnote{The Chinese Hard X-ray Modulation Telescope, NASA's
Nuclear Spectroscopic Telescope Array (NuSTAR), and the Indian X-ray
satellite, ASTROSAT, are possible examples.} may provide additional constraints.

\subsection{Disks and Nuclei}
\label{subsec:disks}
Radio continuum emission is a tracer of massive star formation and does not suffer from
obscuration as do many other tracers such as H$\alpha$ emission.  This is particularly
important for galaxies that are edge-on to the line of sight and so have high optical
depths at other frequencies. 
For example, the H$\alpha$ SFR, corrected for obscuration using $\lambda 24~\mu$m data
\citep{cal07} has not been well-calibrated for edge-on galaxies.
 With corrections for contributions from lower mass stars
\citep[e.g.][]{irw11}, the global radio emission in the disks of the CHANG-ES
galaxies should therefore provide a good measure of total SFR which can be compared to
SFRs from other tracers.   The radio spectral index maps can also provide information
on locations of star formation and CR transport within the disk (see Sect.~\ref{subsec:cr-transport}).

Compact radio cores are indicative of the presence of AGNs
and we also know that some
spiral galaxies exhibit nuclear jets and/or outflows \citep[e.g.][]{hum83,wil83,irw88}.
Galaxy bulges are believed to be correlated with central black holes
\citep{geb00,fer00}
but bulgeless systems can also harbour AGNs \citep{sat09}.  
Although we have some indication as to whether nuclear activity is present in our galaxy
sample (see Table~\ref{table:sample}, `Nuclear Type'),
the presence of aligned emission
or compact radio cores should help to distinguish whether nuclear activity is 
present. 
 These results can be compared to information at other wavebands and can also
be correlated with galaxy environment.



\section{The Galaxy Sample}
\label{sec:sample}
The
sample was taken from the Nearby Galaxies Catalog \citep[NBGC,][]{tul88} which lists
galaxy parameters in a consistent fashion and also provides other useful 
data such
as a parameter indicating local galaxy density.
 The
criteria for including a galaxy in our sample were:\hb
{\it a)} Inclination,  $i\,>\,75^\circ$, to ensure a nearly edge-on aspect,\hb
{\it b)} Declinations, $\delta\,>\,-23^\circ$, for 
accessibility (with adequate uv coverage) from the EVLA,\hb
{\it c)} Blue isophotal diameters, $4\,<\,$d$_{25}~({\rm arcmin})\,<\,15$;  the lower limit
ensures sufficient spatial resolution and the upper limit avoids large galaxies
that would require many pointings and mosaicing.\hb
{\it d)} Flux densities at 1.4 GHz, S$_{1.4}\,\ge\,23$ mJy, to ensure the likelihood of
a detection.

In addition, 3 galaxies which fell just outside of the above
limits were also added because they were known to have extraplanar emission and for which good
ancillary data were available.  These were NGC~5775 (d$_{25}\,<\,4$ arcmin), 
 NGC~4565 (d$_{25}\,>\,15$ arcmin), and
NGC~4244 (d$_{25}\,>\,15$ arcmin and S$_{1.4}\,<\,23$ mJy). There are 35 galaxies in total.

Table~\ref{table:sample} illustrates the range of properties of the galaxies. The distances range from
3.3 to 42 Mpc, 
the galaxies' morphological types span the range of spirals, both barred and unbarred and 
a variety of nuclear
activity types are represented.  They are also in a
 variety of environments (see $\rho$ in Table~\ref{table:sample}), 
from apparently isolated galaxies (e.g. NGC~2683) to
 Virgo Cluster members.  A very nice aspect 
of this survey is that, other than the 3 galaxies
mentioned above, we have not specifically targetted galaxies with known halos and, as a result, there are some
galaxies in the sample which have not previously been searched for halo emission.

The choice of systems with detectable radio flux
 suggests that our galaxies are actively forming stars, although contributions from radio cores 
are also present in some cases.  Star formation rates (SFRs, Table~\ref{table:sample}), however, 
are modest, the maximum being 20 M$_\odot$ yr$^{-1}$ (NGC~4666) or, among those with
exclusively HII region nuclear spectra (i.e. no known compact nuclear activity)
the maximum is
6.4 M$_\odot$ yr$^{-1}$ (NGC~5792).  
By comparison, a sample of 417 nearby star forming galaxies
\citep{mou06} has SFRs ranging from 0.03 to approximately 170 M$_\odot$ yr$^{-1}$ \citep{ken08}.  
Thus, our choice of galaxies with radio continuum fluxes above 23 mJy has not imposed a bias to high
SFR in our sample.   We are probing the halos of `normal' galaxies.



\section{EVLA Observations}
\label{sec:observations}
CHANG-ES has been awarded up to 405 hours of observing time in dynamic scheduling mode at
the EVLA.  
Observations are being carried out at 1.5 GHz and 6 GHz with 500 MHz and 2 GHz bandwidths, 
respectively\footnote{Radio
frequency interference may reduce these values somewhat in practice
(see Paper II).}.  EVLA continuum observations are now made in spectral line mode which
facilitates radio frequency interference (RFI)
 mitigation and also allows for multi-frequency synthesis techniques
to be applied during imaging
(see Paper II).  We are using 2048 spectral channels at 1.5 GHz and
1024 channels at 6 GHz\footnote{Hanning smoothing will effectively
set the
spectral resolution at twice the channel width.} which also permits the effective
application of RM synthesis
techniques (Sect.~\ref{subsec:faraday}).

The distribution of time over frequencies and arrays is
given in Table~\ref{table:obs_time}.  These values are best-case estimates, given that the observing
time has been awarded in `Shared Risk' mode\footnote{Time has been granted
under the `Resident Shared Risk Observing' (RSRO) program
({\tt https://science.nrao.edu/facilities/evla/early-science/rsro}).} during the extensive
commissioning period of the EVLA. All polarization products (Stokes I, Q, U, and V)
are being obtained.

Observations in the B, C and D array configurations at 1.5 GHz and in
C and D configurations at 6 GHz ensure that a variety of spatial scales are probed,
as outlined in Sect.~\ref{sec:motivation}. 
Table~\ref{table:obs_time} shows that the linear resolution spans a range from 56 pc
to 9.4 kpc, depending on galaxy distance, observing frequency and array configuration.  
In fact, the new wide bandwidths provide
even more flexibility than the table implies since 
the resolution will vary between the upper and lower ends of each observing band.  With 
appropriate uv weightings, the array can also be `tuned' to a wider range of spatial resolutions.

The {\it highest} linear resolutions range from 69 pc to
0.88 kpc at 1.5 GHz and 56 pc to 0.71 kpc at 6 GHz, depending on galaxy distance.  For science
which requires a comparison between galaxies at similar {\it linear} resolutions, we will either
spatially smooth the data, apply appropriate models, and/or limit the number of comparison galaxies, 
as necessary.  An example is a comparison of vertical scale heights between galaxies.  A spatial resolution
of approximately 500 pc should permit the measurement of a scale height, assuming an exponential
decline.  At 1.5 GHz, this is possible for
all galaxies closer than 24 Mpc (25 out of 35 galaxies) and at 6 GHz, galaxies closer than 30 Mpc
(30 out of 35 galaxies) meet this criterion.  Note, however, that we may detect emission
to much greater distances; for example, the $\lambda\,$20 cm map of \citet{hum90}
shows the spectacular radio continuum
halo of NGC~4631 being traced to
at least 10 kpc from the plane.


On the other hand, the largest angular size detectable by the array is fixed by the
shortest antenna spacing.  At 1.5 GHz, this is 16 arcmin which is comparable to the optical
diameter of our largest galaxies (Table~\ref{table:sample}), but at 6 GHz, the largest detectable
angular size is only 4 arcmin.  Therefore, missing flux must be supplied
by supplementary single-dish observations of the largest galaxies at 1.5 GHz and of all galaxies
at 6 GHz\footnote{At the time of writing, we anticipate that
these data may come from the
Greenbank Telescope, which will have wide bandwidth receivers installed in 2012.}.
This is especially important in order to derive accurate spectral index
maps between 1.5 GHz and 6 GHz and also to derive proper scaleheights
at 6 GHz. Details as to 
how the single dish data will be combined with the EVLA data will be described in a future paper. 

At 1.5 GHz, the
full width at half maximum of the primary beam ($\theta_{PB}$) is 30 arcmin so only
one pointing is necessary for all galaxies.  At 6 GHz, however, $\theta_{PB}$ = 7.5 arcmin.
Considering the sensitivity of the array to emission outside of the primary beam as well as
observing time requirements, we have opted to observe all galaxies with d$_{25}\,>\,1.3\,\theta_{PB}$ 
with two pointings.  These pointings are placed on the major axis on either side of the
nucleus, separated by one-half of the primary beam.

The noise limits (Table~\ref{table:obs_time}) improve over previous observations
at comparable spatial resolution. Note that the estimated confusion limit applies to
the total intensity image and will be lower for Q and U images.
In addition, it should be possible to make corrections for confusion in 
 the low resolution data (e.g. D array)
using the high resolution (B array) results. 


A description of the data reduction procedures for EVLA data are provided in Paper II.

\section{On the Separation of Thermal and Non-thermal Emission}
\label{sec:separation}

Globally, we know that radio continuum emission in galaxies is dominated by
the non-thermal component at our observing frequencies.  However, the thermal
component is not negligible; for example, 
 thermal contributions are typically 12\% and 26\% of the total flux at 1.5 GHz
and 6 GHz, respectively \citep{con92}.  Since the bulk of the global
emission is from the galaxy's disk rather than the halo, these fractions should
also be typical of disks.  
Few measurements of spectral index have so far been made in fainter galaxy halos;
however, in
NGC~4631, halo emission shows
a steeper spectral index than the disk \citep{hum90},
 implying that the thermal contribution is weaker and possibly negligible
in halos.

The highest linear spatial resolutions for CHANG-ES data ($\approx\,60$ pc, 
Table~\ref{table:obs_time}) are sufficient to resolve larger HII regions
in normal galaxies
\citep[for example, the largest HII regions in M~31 exceed 100 pc][]{azi11}.
Therefore, the thermal fraction could be significant and, indeed, even dominant
in some beams in the disk and possibly the disk-halo interface.
In Paper II, for example, we show that the simplest explanation for the
spectral curvature that is observed in the disk of
NGC~4631 at 1.5 GHz is a contribution from thermal emission; we also show that
two star forming complexes have flat spectral indices, consistent with thermal
emission. 

Wideband EVLA observations (together with single-dish data to fill
in missing spacings) contain a wealth of information that can help with the thermal/non-thermal
problem.
With multi-channel observations in two frequency bands, it is possible to obtain at least 3 
spectral index measurements, at least one within each band, and one between the two bands
at every point.  
 Since a thermal spectrum is characterized by
a known spectral index ($I_\nu\,\propto\,\nu^{-0.1}$) which does not vary with
frequency, a measurement of 3 spectral indices may be sufficient to separate thermal from
non-thermal emission in the case of a constant non-thermal spectral index, $\alpha_{NT}$.
New algorithms, however, are permitting the measurement of changing spectral indices
{\it within} observing bands (see Paper II for details).  For positions with sufficient
signal-to-noise, curvature in $\alpha_{NT}$ can be detected.  This is
particularly important for modelling cosmic ray propagation (e.g. Sect.~\ref{subsec:cr-transport}).

\section{Summary}
\label{sec:conclusions}

In this paper, we have introduced a new survey, called CHANG-ES,
to observe radio continuum halos in 35 edge-on, normal spiral galaxies. The galaxies
are being observed in all polarization products in two different bands, 1.5 GHz (L-band) and 6 GHz
(C-band), and over 3 different EVLA array configurations.  This is the first comprehensive
radio continuum survey of halos to include all polarization products.

Our scientific goals include understanding the physical conditions and origin of 
gaseous halos, probing cosmic ray transport, and determining outflow speeds and
the ultimate fate of the gas.  Mapping
the halo magnetic field configuration and investigating how fields couple to
outflows are particularly important goals, as is 
examining the FIR-radio continuum relation in halos. 
We will also measure
synchrotron scale heights and attempt to reconcile these with models that focus on high energy 
(including gamma-ray) emission.  Ancillary data, such as infra-red and
X-ray data will be applied to some of these questions.

This survey exploits the new capabilities
of the EVLA, especially the wide bandwidths that contribute to
reducing on-source integration times in the detection of galaxy halos.  With up to 405
hours of observing time, our combination of number of galaxies observed,
range of spatial scales probed, and low noise levels 
(see Tables~\ref{table:sample} and \ref{table:obs_time})
significantly improve upon previous
surveys of edge-on galaxies carried out at these two frequencies; in the case of
noise levels, the improvement is more than an order of magnitude.  Spectral index variations
can also be observed and used to separate thermal from non-thermal emission as well as to
model non-thermal spectral index variations in regions of sufficiently high signal-to-noise.

Our first CHANG-ES test observations of NGC~4631 at C array, along with details as to the data reduction
procedures, are reported in Paper II.






\acknowledgments

JAI and DJS would like to thank the staff at the EVLA for their warm welcome and assistance
during their sojourn in Socorro.  Research at Ruhr-Universit{\"a}t, Bochum, is supported by 
Deutsche Forschungsgemeinschaft
 through grants, FOR1048 and FOR1254.
The Digitized Sky Surveys were produced at the Space Telescope Science Institute under 
U.S. Government grant NAG W-2166. The Second Palomar Observatory Sky Survey (POSS-II) 
was made by the California Institute of Technology with funds from the National Science 
Foundation, the National Geographic Society, the Sloan Foundation, the Samuel 
Oschin Foundation, and the Eastman Kodak Corporation. 
The National Radio Astronomy Observatory is a facility of the National
Science Foundation operated under cooperative agreement by Associated
Universities, Inc.



{\it Facilities:} \facility{EVLA} 

\clearpage




\clearpage

\clearpage

\begin{deluxetable}{cccccccccccccc}
\tabletypesize{\scriptsize}
\rotate
\renewcommand{\arraystretch}{0.90}
\setlength{\tabcolsep}{0.05in} 
\tablecaption{The CHANG-ES Galaxy Sample\label{table:sample}}
\tablewidth{0pt}
\tablehead{
\colhead{Name} & \colhead{RA (J2000)\tablenotemark{a}} & 
\colhead{DEC (J2000)\tablenotemark{a}} &
\colhead{i\tablenotemark{b}}
& \colhead{d$_{25}$\tablenotemark{c}} & \colhead{$V_{\odot}$\tablenotemark{d}} &
\colhead{D\tablenotemark{e}} & \colhead{Type\tablenotemark{f}} & \colhead{Nuclear} &
\colhead{ S$_{1.4}$ (ref)\tablenotemark{g}} & \colhead{L$_{FIR}$\tablenotemark{h}} &
\colhead{SFR\tablenotemark{i}} &
\colhead{log(M$_{T}$)\tablenotemark{j}} & \colhead{$\rho$\tablenotemark{k}} \\
\colhead{} & \colhead{(h m s)} & \colhead{(d m s)} & (deg) &
\colhead{($^\prime$) } & \colhead{(km/s)} &
\colhead{(Mpc)} & \colhead{} & \colhead{Type\tablenotemark{a}} & 
\colhead{(mJy)} & \colhead{(10$^{9}$ L$_\odot$)} &
\colhead{(M$_\odot$ yr$^{-1}$)} &
\colhead{(log(M$_\odot$))} & \colhead{(Mpc$^{-3}$)}
}
\startdata
		N~~660 & 01h43m02.40s & +13d38m42.2s &77 & 7.2 & 856  & 12.3 & SBa & HII,LINER & 373 (A)   & 16.8 & 5.7  & 10.73 & 0.12 \\
		N~~891 & 02h22m33.41s & +42d20m56.9s &84 &12.2 & 529  & 9.48 & Sb & HII & 286 (C)          & 12.1 & 4.7 & 11.19 & 0.55 \\
		N 2613 & 08h33m22.84s & -22d58m25.2s &85 &6.8  & 1679 & 23.4 & Sb & HII & 69 (B)           & 9.68 & 4.6 & 11.58 & 0.15 \\
		N 2683 & 08h52m41.35s & +33d25m18.5s &79 &9.1  & 415  & 6.27 & Sb  & LINER,Sy2 & 84 (B)    & 0.86 & 0.36 & 10.81 & 0.09 \\
		N 2820 & 09h21m45.58s & +64d15m28.6s &90 &4.1  & 1576 & 26.5 & SBc &  & 46.6 (A)           & 5.91 & 2.0 &10.87 & 0.28 \\
		N 2992 & 09h45m42.00s & -14d19m35.0s &90 &4.4  & 2212 & 34.0 & Sa & Sy1,Sy2 & 226 (A)      & 16.6 & 7.2 &11.24  & 0.21 \\
		N 3003 & 09h48m36.05s & +33d25m17.4s &90 &6    & 1480 & 25.4 & Sbc &  & 30.4 (A)           & 3.87 & 1.5 &10.75 & 0.64 \\
		N 3044 & 09h53m40.88s & +01d34m46.7s &90 &4.4  & 1335 & 20.3 & SBc & HII & 103 (A)         & 7.14 & 2.6 &10.81 & 0.19 \\
		N 3079 & 10h01m57.80s & +55d40m47.3s &88 &7.7  & 1125 & 20.6 & SBcd& LINER,Sy2 & 821 (C)   & 39.1 & 13 &11.27 & 0.29 \\
		N 3432 & 10h52m31.13s & +36d37m07.6s &82 &4.9  & 611  & 9.42 & SBm & LINER,HII & 120 (B)   & 1.31 & 0.39 &10.19 & 0.19 \\
		N 3448 & 10h54m39.24s & +54d18m18.8s &78 &4.9  & 1391 & 24.5 & S0-a &  & 51.3 (A)          & 6.65 & 2.1 &10.95  & 0.24 \\
		N 3556 & 11h11m30.97s & +55d40m26.8s &81 &7.8  & 697  & 14.09 & SBc & HII & 217 (A)        & 12.5 & 4.9 &10.78 & 0.15 \\
		N 3628 & 11h20m17.01s & +13d35m22.9s &87 &14.8 & 846  & 8.50  & Sb  & HII,LINER & 470 (C)  & 6.99 & 2.3 &11.17 & 0.39 \\
		N 3735 & 11h35m57.30s & +70d32m08.1s &85 &4    & 2696 & 42.0  & Sc & Sy2 & 85.8 (A)        & 24.3 & 11 &11.38 & 0.19 \\
		N 3877 & 11h46m07.80s & +47d29m41.2s &85 &5.1  & 894  & 17.7    & Sc & HII & 40.7 (A)      & 5.19 & 2.2 &10.74 & 1.53 \\
		N 4013 & 11h58m31.38s & +43d56m47.7s &84 &4.7  & 835  & 16.0    & Sb & HII,LINER & 38.2 (A)& 4.25 & 1.8 &10.79 & 1.34 \\
		N 4096 & 12h06m01.13s & +47d28m42.4s &82 &6.4  & 577  & 10.32  & SABc & HII & 51.4 (A)      & 1.77 & 0.75 &10.59 & 0.40 \\
		N 4157 & 12h11m04.37s & +50d29m04.8s &90 &7    & 771  & 15.6   & SABb & HII & 185 (A)       & 9.18 & 3.8 &11.00 & 1.19 \\
		N 4192 & 12h13m48.29s & +14d54m01.2s &83 &8.7  & -142 & 13.55 (V) & SABb & HII,Sy,LINER & 23.0 (A) & 3.3 & 1.49 &11.15 & 1.44 \\
		N 4217 & 12h15m50.90s & +47d05m30.4s &78 &5.1  & 1028 & 20.6   & Sb & HII & 120 (A)         & 11.8 & 5.3 &11.03 & 0.95 \\
		N 4244 & 12h17m29.66s & +37d48m25.6s &90 &15.8 & 247  & 3.30   & Sc & HII & 20.3 (C)        & $<$0.03& $<$ 0.01 &10.02 & 0.39 \\
		N 4302 & 12h21m42.48s & +14d35m53.9s &89 &4.7  & 1118 & 19.41 (V) & Sc & Sy,LINER & 34.8 (A) & 5.08& 2.3 &10.90 & 3.60 \\
		N 4388 & 12h25m46.75s & +12d39m43.5s &79 &5.6  & 2607 & 16.60 (V) & Sb & Sy2 & 119 (A)      & 4.71 & 2.6 &10.94 & 1.56 \\
		N 4438 & 12h27m45.59s & +13d00m31.8s &78 &9.1  & 259  & 10.39 (V) & Sa & LINER & 63.3 (A)   & 0.90 & 0.32 &10.91  & 2.67 \\
		N 4565 & 12h36m20.78s & +25d59m15.6s &90 &16.2 & 1228 & 27.1 & Sbc & Sy3,Sy1.9 & 134 (D)    & 15.7 & 7.9 &11.80 & 1.00 \\
                N 4594 & 12h39m59.43s & -11d37m23.0s &79 &8.4  & 1127 & 12.7 & Sa & LINER,Sy1.9 &  93.4 (A) & 1.59 & 0.68 &11.08  & 0.32 \\
		N 4631 & 12h42m08.01s & +32d32m29.4s &85 &14.7 & 613  & 7.55 & SBcd &  & 772 (C)            & 8.49 & 2.9 &10.74 & 0.41 \\
		N 4666 & 12h45m08.59s & -00d27m42.8s &76 &4.2  & 1516 & 27.5 & SABc & HII,LINER & 434 (A)   & 53.8 & 20 &11.13 & 0.54 \\
		N 4845 & 12h58m01.19s & +01d34m33.0s &81 &4.8  & 1228 & 16.98 (V) & Sab & HII,LINER & 43.3 (A)& 5.17 & 1.8 &11.32 & 0.49 \\
		N 5084 & 13h20m16.92s & -21d49m39.3s &90 &12.5 & 1728 & 23.4 & S0 & poss.LINER & 45.9 (A)   & 0.49 & $<$ 0.21 &11.91 & 0.29 \\
		N 5297 & 13h46m23.68s & +43d52m20.5s &89 &5.3  & 2404 & 40.4 & SABc &  & 23.0 (A)           & 8.09 & 3.9 &11.28 & 0.79 \\
		N 5775 & 14h53m58.00s & +03d32m40.1s &84 &3.9  & 1582 & 28.9 & Sbc &  & 280 (A)             & 38.1 & 14 &11.09 & 0.67 \\
		N 5792 & 14h58m22.71s & -01d05m27.9s &81 &7.2  & 1930 & 31.7 & SBb & HII & 51.8 (A)         & 17.2 & 6.4 &11.47 & 0.52 \\
		N 5907 & 15h15m53.77s & +56d19m43.6s &90 &11.2 & 666  & 15.26 & Sc & HII & 50.5 (A)         & 5.56 & 2.7 &11.32 & 0.26 \\
		U10288 & 16h14m24.80s & -00d12m27.1s &90 &4.9  & 2046 & 34.1 & Sc &  & 26.1 (A)             & 2.55 & 1.3 &11.07 & 0.23 \\
		\hline
\enddata
\tablecomments{Data are from \citet{tul88} unless otherwise indicated.}
\tablenotetext{a}{From the NASA Extragalactic Database (NED).}
\tablenotetext{b}{Inclination, $i\,=\,3deg + cos^{-1}(\sqrt{((b/a)^2 - 0.2^2)/(1 - 0.2^2))}$, where $b/a$ is the minor
to major axis ratio.}
\tablenotetext{c}{Observed blue diameter at the 25th mag/arcsec$^2$ isophote.}
\tablenotetext{d}{Heliocentric radial velocity.}
\tablenotetext{e}{Distance from NED, assuming $H_0\,=\,73$ km s$^{-1}$ Mpc$^{-1}$ and correcting for Virgo
Cluster and Great Attractor perturbations, except for Virgo Cluster galaxies which have been taken from
\citep{sol02}; the latter are indicated by (V).}
\tablenotetext{f}{From HYPERLEDA, {\tt http://leda.univ-lyon1.fr/}, see also \citet{pat03}.}
\tablenotetext{g}{Flux density at 1.4 GHz from refs: (A) NRAO VLA Sky Survey \citep[NVSS][]{con98},
(B) \citet{irw99}, (C)  \citet{str04}, (D) \citet{con02}.
}
\tablenotetext{h}{FIR luminosity from L$_{FIR}$ = 4$\pi$ D$^2$F$_{FIR}$, where
F$_{FIR}$ = $1.26\,\times\,10^{-14}\left(2.58\,S_{60}\,+\,S_{100}\right)$ W m$^{-2}$ for
$S$ in Jy \citep[see][]{san96}, representing the flux in the 40.2 to 122.5 $\mu$m range.  
We take $S$ from \citet{san03} or, if not available, from the most recent entry in NED,
usually the IRAS Faint Source Catalog (FSC).
N~4302 had no NED entry and was taken from the FSC directly.
}
\tablenotetext{i}{SFR(M$_\odot$ yr$^{-1}$) = $4.5\,\times\,10^{-44}$ L$_{TIR}$(ergs s$^{-1}$) \citep{ken98}
where L$_{TIR}$ (3 to 1100 $\mu$m) is computed from $S_{25}$,  $S_{60}$ and $S_{100}$ according to the prescription
in \citet{dal02} and the flux densities are from the same sources as in
the previous note. These values may be refined in the future with improved IR data \citep[see, e.g.][]{mur08}.
}
\tablenotetext{j}{(Log of) the total mass in units of M$_\odot$, where  M$_T\,=\,W^2\,d/(8G)$.
$W$ is the HI line width corrected for random gas motions and galaxy inclination, and $d$ is the
galaxy's linear diameter calculated from d$_{25}$ and the galaxy's distance, corrected for the effects of projection
and obscuration (values adjusted to our distance).}
\tablenotetext{k}{Density of galaxies brighter than -16 mag in the vicinity of the galaxy.
The local density was determined on a 3D-grid at 0.5 Mpc spacing.  No correction has been made
for the fact that our distances differ somewhat from those of \cite{tul88}. }
\end{deluxetable}

\clearpage

\begin{deluxetable}{ccccccccccccc}
\tabletypesize{\scriptsize}
\renewcommand{\arraystretch}{0.90}
\tablecaption{Estimated Observing Time, rms \& spatial coverage for CHANG-ES\label{table:obs_time}}
\tablewidth{0pt}
\tablehead{
\colhead{Array:  $\nu_0$\tablenotemark{a}} & \colhead{Time/pt.\tablenotemark{b}} & 
\colhead{Theor. rms\tablenotemark{c}} & \colhead{Confusion\tablenotemark{d}} & 
\colhead{$\theta_{HPBW}$\tablenotemark{e}} &\colhead{Lin. Res.\tablenotemark{f}}
& \colhead{$\theta_{LAS}$\tablenotemark{g}}\\
\colhead{} & \colhead{} & 
 \colhead{($\mu$Jy beam$^{-1}$)} & \colhead{($\mu$Jy beam$^{-1}$)} & \colhead{(arcsec)} & \colhead{Min/max (pc/kpc)}
& \colhead{(arcmin)}\\
}
\startdata
B:~~~1.5 GHz & 2 hrs & 7.7 & -- & 4.3 & 69/0.88 & 2.0\\
C:~~~1.5 GHz & 30 min &  16 & 9 & 14 & 224/2.85 & 16.2\\
~~~~~~~~~6 GHz & 3 hrs & 2.4 & 0.2 & 3.5 & 56/0.71 & 4.0\\
D:~~~1.5 GHz & 20 min & 20 & 89 & 46 & 736/9.4 & 16.2\\
~~~~~~~~~6 GHz & 40 min & 4.8 & 2.3 &12 & 192/2.4 & 4.0\\
\enddata
\tablecomments{These estimates represent a best-case scenario in which there is no RFI or other errors.}
\tablenotetext{a}{EVLA array configuration and central frequency.}
\tablenotetext{b}{On-source time per pointing, not including overheads. 
At 6 GHz, the field of view is smaller and the larger galaxies require more than one pointing.}
\tablenotetext{c}{Theoretical noise, ignoring confusion, for each polarization product. 
Robust weighting \citep{bri95} 
is assumed.  At 6 GHz, the bandwidth, B = 2 GHz and at 1.5 GHz, B = 500 MHz.
See the EVLA Observational Status Summary (OSS) at {\tt http://evlaguides.nrao.edu/index.php?title=Category:Status}.}
\tablenotetext{d}{Estimated confusion limits (from the OSS).  These confusion limits should only apply to
total intensity, I, but not to polarization products, Q, U or V.}
\tablenotetext{e}{Spatial resolution, assuming uniform weighting. Actual values will be larger when
robust weighting is used. Given the wide frequency bands and possibility
of a variety of uv weightings, the range of spatial resolution is actually greater than noted here.}
\tablenotetext{f}{Minimum and maximum linear resolution for minimum and
maximum galaxy distances, respectively, corresponding to the spatial resolutions of the previous column.}
\tablenotetext{g}{$ $Largest angular size scale detectable by the array.}
\end{deluxetable}





\end{document}